%% file: main.tex
\documentclass[conference]{IEEEtran}
\usepackage{cite}
\usepackage{amsmath,amssymb,amsfonts}
\usepackage{algorithmic}
\usepackage{graphicx}
\usepackage{textcomp}
\usepackage{xcolor}
\usepackage{url}
\usepackage{subfiles}
\usepackage{msc}
\usepackage{listings}

\newcommand{\pub}[1]{\textit{pub}^{#1}}
\newcommand{\sign}[2]{\textit{Sign}_{\textit{prv}^{#1}}(#2)}
\newcommand{\sigpod}[2]{\textit{sig}^{\textit{#1}}_{#2}}
\newcommand{\pk}{\textit{pk}} % zkSNARKs proving key
\newcommand{\vk}{\textit{vk}} % zkSNARKs verifying key
\newcommand{\pkd}{\textit{pk}^D} % zkSNARKs proving key for delivery
\newcommand{\vkd}{\textit{vk}^D} % zkSNARKs verifying key for deliver
\newcommand{\pke}{\textit{pk}^E} % zkSNARKs proving key for exchange
\newcommand{\vke}{\textit{vk}^E} % zkSNARKs verifying key for exchange

\def\BibTeX{{\rm B\kern-.05em{\sc i\kern-.025em b}\kern-.08em
    T\kern-.1667em\lower.7ex\hbox{E}\kern-.125emX}}

\begin{document}

\title{Towards Decentralized IoT Updates Delivery Leveraging Blockchain and Zero-Knowledge Proofs}

\author{\IEEEauthorblockN{Edoardo Puggioni\IEEEauthorrefmark{1}\IEEEauthorrefmark{2}, Arash Shaghaghi\IEEEauthorrefmark{1}, Robin Doss\IEEEauthorrefmark{1}, and Salil S. Kanhere\IEEEauthorrefmark{3}}
\IEEEauthorblockA{\IEEEauthorrefmark{1}Deakin University, Geelong, Australia -- Centre for Cyber Security Research and Innovation (CSRI)\\
\IEEEauthorrefmark{2}Dipartimento di Automatica e Informatica, Politecnico di Torino, Italy \\
\IEEEauthorrefmark{3}The University of New South Wales (UNSW Sydney), Australia\\
\{epuggioni, a.shaghaghi, robin.doss\}@deakin.edu.au\\
salil.kanhere@unsw.edu.au}
}

\maketitle

\begin{abstract}
Internet of Things (IoT) devices are being deployed in huge numbers around the world, and often present serious vulnerabilities. Accordingly, delivering regular software updates is critical to secure IoT devices. Manufactures face two predominant challenges in providing software updates to IoT devices: 1) scalability of the current client-server model and 2) integrity of the distributed updates - exacerbated due to the devices' computing power and lightweight cryptographic primitives. Motivated by these limitations, we propose \textit{CrowdPatching}, a blockchain-based decentralized protocol, allowing manufacturers to delegate the delivery of software updates to self-interested distributors in exchange for cryptocurrency. Manufacturers announce updates by deploying a smart contract (SC), which in turn will issue cryptocurrency payments to any distributor who provides an unforgeable proof-of-delivery. The latter is provided by IoT devices authorizing the SC to issue payment to a distributor when the required conditions are met. These conditions include the requirement for a distributor to generate a zero-knowledge proof, generated with a novel proving system called zk-SNARKs. Compared with related work, CrowdPatching protocol offers three main advantages. First, the number of distributors can scale indefinitely by enabling the addition of new distributors at any time after the initial distribution by manufacturers (i.e., redistribution among the distributor network). The latter is not possible in existing protocols and is not account for. Secondly, we leverage the recent common integration of gateway or Hub in IoT deployments in our protocol to make CrowdPatching feasible even for the more constraint IoT devices. Thirdly, the trustworthiness of distributors is considered in our protocol, rewarding the honest distributors' engagements. We provide both informal and formal security analysis of CrowdPatching using Tamarin Prover.
\end{abstract}

\begin{IEEEkeywords} Internet of Things (IoT), Secure Update Delivery, Blockchain, Smart Contracts, zk-SNARKs, Tamarin Prover \end{IEEEkeywords}

\section{Introduction} \label{intro}
The number of Internet of Thing (IoT) devices deployed worldwide is growing at an incredible speed projected to reach more than 75 billion by 2025 \cite{statistica}. Security is often an afterthought for the manufacturers of these devices, with their primary concern being cost, size, and usability of devices. Hence, IoT devices often present vulnerabilities that attackers can exploit, posing serious threats to organizations and individuals' security and privacy (e.g., IoT devices turned into zombie-agents to perpetrate DDoS attacks). Accordingly, regular software updates are fundamental in securing vulnerable IoT devices and present a number of key challenges \cite{lin_iot_2016,kimani_cyber_2019,acostapadilla_iotfuture_2016, hernandez-ramos_updating_2020}. 

One of the most important requirements is integrity of the updates. It is fundamental that updates are not tampered with, otherwise the patching process would create more threats than it mitigates, leading to malicious code being executed.
% Another challenge is trust. That is, IoT devices must be able to trust software providers, to avoid illegitimate software being released from unauthorized entities. In particular, this can be achieved through the use of proper cryptographic tools and key management algorithms.
Secondly, IoT objects are usually characterized by restricted hardware and software resources. Hence, the need for efficient and lightweight cryptographic primitives and protocols. Thirdly, given the number of devices involved, another crucial aspect is scalability. Manufacturers currently deliver updates by means of traditional client-server architectures. However, this approach is clearly not sustainable when we consider the number of devices, different versions of the same devices, iteration of updates, and requirement to support legacy devices.

In this paper, we propose \textit{CrowdPatching}, a protocol allowing manufacturers to deliver updates to IoT devices in a decentralized manner, leveraging blockchain (BC) technologies and zero-knowledge proofs. In this system, the manufacturer delegates the delivery of new updates to self-interested agents, called distributors, who undertake this task in exchange for cryptocurrency (CC) payments. First of all, we assume to have a permissionless BC as an underlying infrastructure, supporting native CC and smart contracts (SCs). Manufacturers announce a new update release by deploying a SC, which in turn will issue CC payments to any distributor who provides a proof-of-delivery.
%The latter is a signature through which each target IoT device is attesting to have received an encrypted version of the update file obtained employing a specific key that has a known hash value. In this context, the use of a zero-knowledge tool, called zk-SNARKs, is fundamentally important. Before a proof-of-delivery can be issued, a distributor is expected to provide a zk-SNARKs proof about said encrypted update, proving that (1) it was indeed obtained encrypting the update file authorized by the manufacturer and (2) the employed key has indeed that known hash value. If this zero-knowledge proof is valid, the proof-of-delivery signature is generated. The latter is sent to the SC by the distributor, along with the encryption key. Before issuing the payment, the contract checks that that the signature is valid, and that the key corresponds to the known hash value. Along with the payment, the SC will publish the key through which the update can be decrypted.
The latter is a signature generated by an IoT device. It works as an unforgeable digital commitment through which an IoT device is authorizing the SC to issue a payment to a certain distributor. For this reason, this important signature is produced by an IoT device only if specific conditions are met. In particular, a distributor is required to provide an encrypted version of the update, along with the hash value of the symmetric key employed. What is more, a distributor needs to produce evidence that this encrypted file was indeed derived from the official update, using a key with the provided hash value.
In this context, the use of a zero-knowledge tool, called zk-SNARKs, is fundamentally important.
Before a proof-of-delivery can be issued, a distributor is expected to provide a zk-SNARKs proof about the provided encrypted update, mathematically proving that (1) it was indeed obtained encrypting the update file authorized by the manufacturer and (2) the employed key has indeed the provided hash value. Most importantly, this zero-knowledge proof does not reveal anything else beyond those facts: the unencrypted update and the key remain secret. If the proof is valid, then the proof-of-delivery signature is generated by the IoT device, and sent to the SC by the distributor along with the encryption key. Before issuing the payment, the contract checks that the signature is valid, and that the key matches the hash value. Along with the payment, the SC publishes the key on the BC. The latter can be now used to decrypt the file, and the update is finally obtained by the IoT device.

Compared to similar proposals \cite{boudguiga_towards_2017, he_securing_2019, lee_patch_2018, leiba_incentivized_2018} discussed in Section \ref{related}, CrowdPatching brings several novelties and optimizations. In particular, we addressed a number of issues in the work of Leiba et al.\ \cite{leiba_incentivized_2018}, which constitutes the main inspiration for our proposal. First of all, our porotocol allows the number of distributors to fully scale with the number of IoT devices. Indeed, any distributor who could not obtain the update file from the manufacturer, can acquire it from other distributors in exchange for a CC payment. This is in contrast with the proposal of Leiba et al.\, where distributors can obtain the update file only in an initial phase, and new distributors willing to participate at a later moment cannot do so in any way. We argue that this limitation significantly hinders the scalability of their system, as it does not allow the number of distributors to grow indefinitely. Furthermore, we drastically reduce the requirements necessary for IoT objects to participate in the protocol, to account for their limited resources. We achieve this by means of a new participant, called hub, which functions as a trusted gateway for IoT devices in their local networks, performing the most demanding protocol steps in place of them. In particular, differently from the work of Leiba et al., IoT devices are not required to participate in the BC network in any way, and can avoid verifying zk-SNARKs proofs. Finally, we introduce a simple mechanism allowing to judge the trustworthiness of distributors, leveraging a data structure automatically maintained by a SC. This mechanism is particularly important in a context where literally anyone can act as a distributor, and no other trust-related assumption can be made about these entities. What is more, we exploit the only reliable information a distributor is required to provide: the public key used as recipient for CC payments. We achieve this through a key-value database stored on the BC, associating each distributor's public key with the number of successful deliveries performed through time. In other words, when a certain distributor delivers an update file for the first time ever, a new entry is added to this database associating its public key with the integer value $1$. After each subsequent successful delivery, this value is incremented. In this way, other participants of the protocol, especially hubs, can access the BC to read this value when choosing a distributor to interact with, selecting the one with the highest score. This mechanism encourages distributors to behave honestly, keeping a secure and unforgeable record of successful deliveries. 

We start this paper by providing the necessary background in Section \ref{back}. We continue discussing related works in Section \ref{related}, where we also point to a security weakness in the protocol proposed in \cite{leiba_incentivized_2018}. In Section \ref{proposal}, we illustrate the details of our proposal, discussing all protocols steps in depth. Furthermore, we provide its security analysis in Section \ref{security}, presenting both an informal analysis and a formal analysis. The latter was performed by means of the Tamarin prover, a state-of-the-art tool for automated protocol security analysis. Finally, we discuss conclusions and future work in Section \ref{concl}.

\section{Background} \label{back}
\subsection{Blockchain and Smart Contracts} \label{blockchain}

A BC is a distributed database constructed as an immutable sequence of blocks. Its purpose is to maintain a permanent record of transactions. Each peer stores a copy of the BC, often referred to as a distributed ledger. Any new block contains the hash value of the previous block. Tampering with any block would break the chain of hash values, provoking a mismatch. All peers agree with the validity of new blocks by means of a consensus protocol. That is, a process in which an active node of the network, called miner, gathers a set of transactions to form a new block and validates it in a way that will be accepted by other peers. The specific protocol depends on the BC implementation.
The BC was employed by Nakamoto in the first cryptocurrency (CC) system, Bitcoin \cite{nakamoto_bitcoin_nodate}, where it was used to solve the double spending problem. All peers agree on the order of transactions, avoiding fraudulent behaviors. The consensus algorithm is called Proof of Work.
%Here, a miner prepares a block and validates it by attaching the solution of a puzzle. A solution which is extremely difficult to find and yet easy to verify. A valid candidate block is then broadcasted to the network, so other peers can verify it and accept it as the next block of the chain.

After Bitcoin, many alternative BC systems were introduced, with applications beyond CC, such as smart contracts. The term was first coined in a paper by \cite{Szabo_1997}, long before the invention of Bitcoin.
%, where the author compared this technology to vending machines.
% Any entity in possession of coins can interact with the machine, which will react automatically providing products (and change) according to the price tag. What is more, security mechanisms are in place, and cost of breaking them is higher than the benefits for a potential malicious user.
However, the first practical applications of such concept were proposed years later, exploiting distributed ledgers and their consensus algorithms. In general, SCs can be defined as computer code deployed on the BC, executed securely and automatically in a distributed fashion when certain conditions are met.
%Similarly to pure CC systems, SCs operations are triggered by valid transactions performed by peers on the network, and the results are also recorded in the distributed ledger.
The most important implementation in this context is Ethereum, introduced by Buterin \cite{buterin_next_2014}.
%and currently the second most widely adopted BC platform after Bitcoin.
Along with SCs, this BC system also supports CC.
%, also named Ethereum.
Ethereum supports externally owned accounts, owned by users of the network, and SC accounts. External accounts are controlled by private keys, SC accounts are controlled by their own code. Users can create a new contract by sending a transaction to a specific fixed address, attaching its code, the needed parameters and an arbitrary amount of CC. Otherwise, users can also (1) exchange CC between each other or (2) send transactions to an address belonging to an existing contract. This last action would trigger the distributed execution of the SC code. Depending on the conditions, many actions can be performed, such us changing the state of the contract by writing in its storage on the BC or automatically issuing transactions to other accounts, which can be users or other SCs.
%To avoid malicious actors exploiting the automatic nature of code execution, all Ethereum operations requiring computational effort have an associated fee, which is measured with a special unit called \textit{gas}.

\subsection{zk-SNARKs} \label{zksnarks}

In general, zero-knowledge proofs systems are interactive protocols in which a prover \textit{P} is intentioned to convince the verifier \textit{V} that a given statement is true, without revealing the secret values included in the statement itself. For example, P could prove the validity of a statement $y = H(x)$, where the secret value $x$ is the hash preimage of a non-secret value $y$, without revealing $x$.
%Traditionally, zero-knowledge proof protocols require the two parties to have an interactive communication composed of several successive steps. 
Zero-Knowledge Succinct Non-interactive Arguments of Knowledge (zk-SNARKs) are a novel zero-knowledge proof system \cite{gennaro_quadratic_2013, bitansky_succinct_2013, ben-sasson_succinct_2013,ben-sasson_snarks_2013} introducing several optimizations. Among them, the fact that they are non-interactive, allowing \textit{P} to generate proofs asynchronously.
%In other words, a proof is generated once and for all, and can be verified at any moment by the verifier.
Furthermore, they are succinct, meaning that (1) the verification process is not computationally expensive and (2) the proofs are limited in size.
The zk-SNARKs system is composed of three algorithms, \textit{Setup}, \textit{Prove} and \textit{Verify}, informally defined as follows:

\begin{enumerate}

    \item The \textit{Setup} is performed by a trusted third party (TTP), who knows the structure of the statement $S$ that needs to be proved (meaning the size of its variables and the algorithms in use) but not the assigned values. Using that structure as input, the proving key $\pk$ and the verifying key $\vk$ are generated. For example, $S$ could be defined as $y = H(x)$, and to generate the keys the TTP would have to know the size of $y$ and $x$, as well as the hashing algorithm in use, e.g. SHA256.
    
    \item Before proceeding, \textit{P} selects valid values for the variables in $S$. Continuing the example, \textit{P} selects $x_1$ and $y_1$ such that $y_1 = H(x_2)$. The \textit{Prove} algorithm takes as input the assigned values, along with the key $\textit{pk}$, and generates the proof $\pi$.

    \item The \textit{Verify} algorithm takes $\pi$, $\vk$ as inputs, along with the non-secret values, i.e. $y_1$ in the example. If the proof is valid, \textit{V} is convinced about the validity of the statement without having acquired any knowledge of its secret values, i.e. $x_1$ in the example.

\end{enumerate}

\section{Related Work} \label{related}

We present here relevant research studies proposing solutions for IoT software updates delivery, targeting the challenges mentioned in Section \ref{intro}.
Firstly, the work of Boudguiga et al.\ \cite{boudguiga_towards_2017} suggests two main strategies employing BC technologies. The first is for the manufacturer to sign an update and upload it to the BC, exploiting its immutability.
% The use of such distributed ledger would provide both scalability and resistance to common availability threats such as DDoS attacks.
However, the authors themselves recognize how uploading the whole update file would easily become unfeasible due to the BC size
% , which must be downloaded by all participants in its entirety. 
Hence, the second strategy, where the role of the BC is limited to integrity purposes. In parallel, the update file is distributed by means of a P2P network, in which IoT objects actively share files with each other. However, no incentive mechanisms is designed to induce IoT nodes to share these files. What is more, IoT resource constraints are not taken into consideration. These devices are often battery-powered and computationally limited, and could not afford an otherwise trivial task such as P2P file sharing.
Similarly, He et al.\ \cite{he_securing_2019} propose the use of a BC to validate an update before installing it. However,
% despite recognizing the importance of not uploading the update file to the BC, 
the authors do not suggest an alternative strategy for its distribution.

A paper by Lee \cite{lee_patch_2018}, presents significant improvements. Scalability is identified as a crucial issue, motivating the proposal of a decentralized system. That is, a framework where the software provider can do away with expensive client-server architectures and the update delivery is delegated to untrusted third parties called transporters. These agents are incentivized through micro-payments in the form of cryptocurrency. However, IoT devices are required to perform many different actions, which are often computationally expensive. The most significant example is the need to function as full BC nodes, which entails monitoring, deploying SCs and managing a wallet. Furthermore, every IoT device receiving a new update file needs to deploy a new SC. Each new deployment will require a new fee, and the cost could easily become overwhelming. Finally, a structural problem prevents the system from scaling in practice. For any given update, the provider is required to share with the transporters a number of update files equivalent to the number of IoT recipients. This would have drawbacks that are comparable to the ones of traditional client-server architectures, e.g. in terms of bandwidth.

Finally, we present the solution proposed by Leiba et al. \cite{leiba_incentivized_2018}, which brings many novelties and optimizations. We will discuss it separately in the next sub-sections, as it constitutes the main inspiration for our own design.

\subsection{Case Study: Protocol}

The proposal of Leiba et al. \cite{leiba_incentivized_2018} consists in a highly decentralized framework allowing any IoT manufacturer to delegate the delivery of software updates to self-interested third-parties.
% The honest behavior of these parties, called \textit{distributors}, is enforced through (1) the use of SCs deployed on a BC and (2) the use of the zk-SNARKs zero-knowledge tool, as well as (3) other traditional cryptographic tools such as public key cryptography and signatures.
First of all, the framework is based on two underlying infrastructures: a permissionless BC, with support for SCs and CC, and a P2P file sharing network.
% Any manufacturer $m$, the entity initiating an instance of the protocol by releasing a new update $U$, is required to participate in both the BC and  the P2P networks. Furthermore, $m$ is required to posses an asymmetric key pair $(\prv{m}, \pub{m})$. Each vendor $m$ is the manufacturer of a specific set of IoT objects and stores list of public keys $\pub{o_{m,i}}$. In addition to its public key, each IoT device $o_{m,i}$ stores its own private key $\prv{o_{m,i}}$, as well as the public key of its manufacturer. Finally, another entity corresponds to the distributors. These are third-party untrusted agents who are offered a cryptocurrency reward in exchange for the completion of a task: deliver the update to IoT devices. They are required to be full participants in both the blockchain network and the P2P network.
In the following, we will briefly illustrate the necessary steps for an instance of the protocol in which a manufacturer aims at delivering an update to a set of IoT objects. Several aspects of this protocol have analogous counterparts in our proposal. See Section \ref{proposal} for an in-depth analysis.

\subsubsection{Update Release} 

An instance of the protocol starts with a manufacturer $m$ releasing an update $U$ for a certain set of IoT devices, identified by their public keys (PKs). Among them is the IoT object $o_m$. The next step for $m$ is to deploy a new SC on the BC, providing a series of elements to initialize its state. The code of the SC corresponds to the following algorithm: for each PK in the list of target IoT objects, if any distributor is able to present a proof-of-delivery (PoD), then a payment will be sent to that distributor as a reward. A certain amount of CC is attached by $m$ at contract creation and will serve as a deposit. The SC will use this deposit to fund the CC rewards.

% Manufacturer $m$ is about to release an update file $U$, computes its hash $U_h := H(U)$ and generates the zk-SNARKs keys $\textit{pk}$ and $\textit{vk}$. Prepares the update package $P$ containing $U$, $\textit{vk}$, $\textit{pk}$ and $\textit{sig}_m$. The latter is a signature made by $m$ on the concatenation of $U$ and $vk$. Formally, $\textit{sig}_m = \sign{m}{U_h || vk}$. The hash of the package $P_h := H(U)$ is also computed. Finally $m$ deploys a new SC by sending several elements to the BC, such as the hash values $U_h$ and $P_h$, and the list of public keys $\pub{o_{m,i}}$
%$\pub{o_{m,1}}, \pub{o_{m,2}}, \dots, \pub{o_{m,n}}$
% belonging to the target IoT objects. Furthermore, $m$ needs to provide the code for the SC, defining its logic. In simple terms, the pseudocode provided by the authors results in the following algorithm. For each public key $\pub{o_{m,i}}$ in the provided list, if any distributor with public key $\pub{d_j}$ is able to present a proof-of-delivery $\textit{sig}_{o_{m,i}}$, then a payment will be sent to that public key $\pub{d_j}$. The nature of the proof-of-delivery $\textit{sig}_{o_{m,i}}$ will be explained later. Finally, $m$ is also required to attach a certain amount of cryptocurrency, which will serve as a deposit. The SC will use this deposit to fund the cryptocurrency rewards.

\subsubsection{Initial Seed}

Distributors become aware of the new release, as they are regularly monitoring the BC. The manufacturer $m$ enters a temporary phase in which $U$ is sent to any distributor who requests it via the P2P network. At the end of this phase, a finite number of distributors has obtained $U$. From this moment on, there is no way for any additional distributor to participate.

\subsubsection{Update Delivery} \label{related:delivery}

Distributors announce the possession of $U$ on the P2P network. IoT objects, who are also regularly monitoring the BC, become aware of the new release and request the update after reading its hash value on the BC. Let us analyze the actions of a distributor $d$ who has received a request from the IoT object $o_m$.
Distributor $d$ sends an identification challenge $c$ to $o_m$, who responds with its PK and a signature on $c$.
Then, $d$ validates the signature and verifies that the PK belongs to the list in the SC. Now $d$ can prepare the zk-SNARKs proof. As explained in Section \ref{zksnarks}, this zero-knowledge proving system allows a party to prove the knowledge of certain elements within a statement to another party, without revealing those secret elements. In this case, $d$ wants to prove three facts to $o_m$. First, the knowledge of a certain key $r$ with hash value $s$, revealing only $s$. Second, the knowledge of a file $U$, revealing only its encrypted version $U_e$ obtained employing $r$ as a symmetric key. Third, the fact that the the hash value of $U$ is $U_h$. The latter is published in the SC by $m$ and can be read by $o_m$. The proof is in the form of a separate file, and can be asynchronously verified by $o_m$ once received from $d$. Along with the proof, $d$ also sends the non-secret values $s$ and $U_e$. If the proof is valid, $o_m$ is mathematically convinced about the three facts above, without gaining any knowledge about $r$ or $U$. Hence, $o_m$ is not yet able to decrypt $U_e$.
% (3) generates a random key $t$; computes (4) $r = H(\pub{d} || t)$ and (5) $s = H(r)$; (6) obtains $U_e = Enc(U, r)$ by applying symmetric encryption to $U$, using $r$ as key; (7) generates the zk-SNARKs proof $\pi$. The zk-SNARKs proof is generated for the following statement $S$, where the public values are $U_h$, $U_e$ and $s$, while the secret values are $U$ and $r$:
% \[ s = H(r) \land U_h = H(U) \land U_e = Enc(U, r) \]
% Finally, $d$ sends a message to $o_m$ containing: the public values $U_h$, $U_e$ and $s$; the proof $\pi$; the signature $\textit{sig}_m$.
% Device $o_m$ receives the message from $d$. Then, it first verifies the signature $\textit{sig}_m$ to authenticate the values $U_h$ and $vk$. Secondly, it verifies the zk-SNARKs proof $\pi$ for the statement $S$. If valid, there will be mathematical assurance about two important facts: (1) the encrypted update $U_e$ received from $d$ was effectively obtained encrypting an update file $U$ with hash value $U_h$ using a key $r$; and (2) the received hash value $s$ was indeed obtained computing the hash of the key $r$.
At this point, $o_m$ can send the PoD to $d$, simply defined as a signature on the concatenation of $U_h$ and $s$. This simple signature has an important role: it can be seen as a formal commitment made by $o_m$, a digital token capable of unlocking the CC reward for $d$.
% This will be clear in the next steps, where the distributor sends the proof-of-delivery to the SC.

\subsubsection{Reward Claim and Key Publication}

Let us continue to follow the actions of $d$ and $o_m$. If the PoD is valid, $d$ posts a transaction to the SC, attaching the PK of $o_m$, the values $t$, $r$ and $s$, and the PoD.
The DSC takes care of validating this transaction in a secure and fair manner, checking that all values were produced correctly and that signatures are valid. If that is the case, a payment is automatically issued to $d$, and the key $r$ is published to the BC. As a consequence, $o_m$ can retrieve the key $r$ and finally decrypt $U_e$ to obtain $U$.

\subsection{Case Study: Discussion}

The work of Leiba et al. \cite{leiba_incentivized_2018} presents several significant improvements compared to other related studies. First of all, IoT objects are not required to be full nodes in the BC network, and can avoid maintaining a CC wallet. Also, only one contract is deployed at the expense of the manufacturer for any single release, regardless of the number of IoT devices. However, during the initial seeding phase, only a finite number of distributors will obtain the update file, and this number cannot increase in any way at a later moment. We argue that this limitation does not allow the framework to scale adequately.
%Indeed, the number of distributors can only grow linearly with the size of the time-window in which the manufacturer seeds the file.
To address this issue in our proposal, we introduce the possibility for distributors to share the update with new distributors in exchange for a CC payment. We argue that this improvement will allow the number of distributors to grow exponentially.

Furthermore, we argue that requirements for IoT nodes are too significant for such resource-constrained devices. Even if they can avoid storing the entire BC data structure, as the authors suggest, they are still expected to access the BC, adding an unnecessary layer of complexity.
Similarly, IoT objects are expected to access the P2P network to interact with distributors, and to verify zk-SNARKs proofs, with analogous consequences.
To overcome this problem, we introduce a new participant in our proposal: the hub. That is, a gateway device managing a heterogeneous set of IoT devices all connected to the same local network. In general, several studies show the importance of such entity in the IoT context, including from a security perspective \cite{cirani_iot_2015, doan_towards_2018, maroof_plar:_2019, simpson_securing_2017}.

Finally, while performing the formal analysis of our protocol, we discovered an important vulnerability that is also applicable to this design. Referring to the steps described in Section \ref{related:delivery}, a malicious distributor could construct the ID challenge as $c := U_h\,||\,s$, asking the IoT device to produce a signature on this value. The malicious distributor can then send this signature to the SC and unlock the cryptocurrency payment without having delivered the update. This effectively breaks the security of the protocol.

\section{Proposed Protocol: CrowdPatching} \label{proposal}

We propose a distributed protocol for IoT updates delivery allowing manufacturers to do away with expensive centralized architectures. We first describe its participants in Section \ref{entities}. Secondly, we illustrate each step in Section \ref{steps}.

\subsection{Protocol Entities} \label{entities}

The protocol includes a P2P file sharing network accessible by anyone, allowing peer discovery by means of a distributed hash table (DHT). A permissionless BC is also required, supporting SCs and a digital currency. What is more, the BC must allow SCs to generate other SCs. An example of such BC platform is Ethereum, described in Section \ref{back}. In the following, we will implicitly refer to the Ethereum implementation as the BC platform in use.

An instance of the protocol is initiated by a manufacturer wanting to release an update for a set of IoT devices. Each of these devices is assumed to posses a secret or private key (SK) and a public key (PK), as well as the PK of the corresponding manufacturer. Furthermore, the protocol assumes IoT objects to be deployed in a local network where they are managed by a hub. That is, a gateway device performing various operations on behalf of the IoT objects for which it is responsible. Hubs are not associated with the manufacturer, and the latter makes no security assumptions about the former. For this reason, we introduce a cryptocurrency (CC) incentive for hubs. On the other hand, hubs are trusted by the corresponding IoT devices. Despite these assumptions, the protocol offers flexibility: any IoT device can undertake both its own role and the one designed for hubs, avoiding the need for a gateway. Additionally, this protocol includes one last entity: distributors. These are self-interested agents whose objective is to obtain CC payments in exchange for delivering updates to IoT devices. We divide them in two categories, first-hand distributors (FHDs) and second-hand distributors (SHDs). FHDs obtain the update file directly from the manufacturer, while SHDs acquire it from FHDs in exchange for a direct CC payment.

\subfile{crowdp-msc.tex}

\subsection{Protocol Steps} \label{steps}

Before any update can be released, the manufacturer is first required to perform a preliminary step and deploy what we call a \textit{Super} SC (SSC) to the BC. That is, a SC which is capable of generating new SCs with a specific template. The generation of these \textit{derived} SCs is triggered when specific transactions are sent to the SSC. This preliminary step is performed only once for each manufacturer. The SSC has an additional purpose: it stores an integer score associated with any distributor. More precisely, an SSC maintains a data structure in the BC keeping track of successful deliveries accomplished by distributors, by means of a simple integer value. If a distributor performs its first delivery of an update to an IoT device, its score is instantiated with value $1$. For any subsequent delivery, its score is incremented. This value, stored on the BC and accessible by anyone, can be used by other participants to judge the relative trustworthiness of any distributor compared to others. Additionally, the score of a distributor is periodically reset to $0$ by the SSC. This mechanism is introduced to avoid a situation in which certain distributors accumulate very high scores, making it very difficult for new distributors (starting with score 0) to be trusted by hubs. The time interval after which the score is periodically set to $0$ is provided by the manufacturer as a parameter for the SSC at its deployment, and it applies to all distributors equally.

Once the SSC is deployed, the manufacturer can proceed. Let us focus on the actions performed by the following actors. A manufacturer $m$ who has already deployed its SSC and is about to release a new update file $U$. One IoT device $o_m$ among the many others manufactured by $m$ and targeted by the release of $U$. The hub $h$ managing $o_m$. One FHD $d_f$ and one SHD $d_s$. The result of the steps described throughout the rest of this section will be threefold: (1) $o_m$ will have received the update file; (2) the distributor responsible for the delivery will have received a CC reward; (3) $h$ will have received a CC reward as well. \figurename \ref{fig:msc1} presents a graphical overview of these steps, without depicting the execution of a DDE (illustrated in Section \ref{seed}) for the sake of clarity.

\subsubsection{Update Release} \label{release}

Firstly, $m$ needs to prepare a series of elements. A pair of zk-SNARKs keys, $\pkd$ and $\vkd$, for the statement $S_D$, where the secret variables are $U$ and $r$:
$$s = H(r) \land U_h = H(U) \land U_e = Enc(U, r)$$
Another pair of zk-SNARKs keys, $\pke$ and $\vke$, for the statement $S_E$, where the secret variables are $P$ and $r$:
$$s = H(r) \land P_h = H(P) \land P_e = Enc(P, r)$$
The purpose of these key pairs will be illustrated in Sections \ref{seed} and \ref{delivery} respectively. To produce them, $m$ only needs the sizes of each variable in the statements (not their actual values) and the specific algorithms used. For example, $r$ could be a 256 bits key, and the hash function $H$ could be implemented with the SHA256 algorithm. $S_D$ has almost the same structure as $S_E$ except for the size of $U$ compared to $P$: for this reason, two different key pairs are needed. The package $P$ is constructed by $m$ to contain $U$, $\pkd$, $\vkd$ and a signature by $m$ on $U_h$, formally defined as $\textit{sig}_{m} = \sign{m}{U_h}$. Now $m$ can send a transaction to the SSC to trigger the creation of a new SC, attaching a CC deposit and several parameters. In particular, the SSC is able to generate two kind of SCs, delivery SCs (DSCs) and exchange SCs (ESCs), depending on which SSC function is addressed by the transaction. In this case, $m$ generates a new DSC. We will discuss ESCs in \ref{seed}. The deposit will serve as a source for CC rewards. The parameters are needed to initialize the state of the newly created SC. It is worth noting that $m$ is not required to attach the contract code, which would be necessary with traditional contract creation. Instead, the code is attached by $m$ only once when the SSC is created. The SSC can then use it to generate other DSCs. This is convenient for $m$, but most importantly for other participants: they can avoid expensive security checks on each new contract code, concentrating on the security of a single SSC instead, and being assured that any DSC derived from it is secure as a consequence. The same reasoning is valid for ESCs. The parameters sent by $m$ are the following: (1) an integer value $e$ indicating the time interval (e.g. in weeks) after which the SC is considered expired; (2) the hash values $U_h := H(U)$, $P_h := H(P)$, $\vkd{_h} := H(\vkd)$, $\vke{_h} := H(\vke)$ and $\pke{_h} := H(\pke)$; (3) the list $L_m$ of PKs belonging to the target IoT objects, containing the key $\pub{o_m}$ among others; (4) the values $a_d$ and $a_h$ representing the amounts of CC for each reward for distributors and hubs respectively.

The SCC performs one check: the selected amounts $a_d$ and $a_h$ must be compatible with the CC deposit sent by $m$ and the number of target IoT objects, to avoid the case in which the SC has not enough CC balance to fund all the rewards. If this check is successful, a new DSC is deployed with a code that translates to the following algorithm. If any distributor with PK $\pub{d}$ is able to present a valid proof-of-delivery (PoD) for any of the IoT targets in $L_m$, and no previous PoD was presented for that target before, then a CC payment (of amount $a_d$) is sent to $\pub{d}$. Furthermore, if any hub with PK $\pub{h}$ provides a valid proof-of-final-delivery (PoFD) for any of the targets, and no previous PoFD was previously presented for that target, a CC payment (of amount $a_h$) is sent to $\pub{h}$. The previous actions can only be performed if the SC is not expired, i.e. if the amount of time passed since its creation is still less than $e$. The nature of PoDs and PoFDs will be explained while illustrating the next steps. In short, they are signatures generated by IoT targets on specific values, and they are able to securely prove that a certain step of the protocol was performed. The DSC can easily verify these signatures using the PKs listed in $L_m$.

\subsubsection{Initial Seed and Additional Sharing} \label{seed}

Distributors are regularly monitoring the BC and become aware of the new release. Initially, $m$ seeds the package $P$ via the P2P network to anyone requesting it through its hash $P_h$ available on the BC. At the end of this temporary phase, a certain number of distributors has obtained $P$. Now, these FHDs compete against each other, trying to be the first to deliver the update to as many IoT devices as possible. Hence, they have no interest in sharing $P$ with any new distributor willing to participate in the protocol. The only way for a SHD to obtain $P$ is to acquire it from a FHD in exchange for a CC payment. Let us follow the actions taken by a SHD $d_s$ to perform such exchange with a FHD $d_f$ who is in possession of $P$. In general, we refer this interaction as a distributor-distributor exchange (DDE).

The SHD $d_s$ sends a request for $P$ on the P2P network, and $d_f$ replies with an identification challenge $c$. To prove its identity, $d_s$ sends back its PK $\pub{d_s}$ along with a signature $sig_{d_s} = \sign{d_s}{c}$. Once the signature is verified, $d_f$ can check the score of $d_s$ on the BC and decide whether to proceed with this interaction. If $d_f$ decides to continue, it generates a zk-SNARK proof for the statement $S_E$. Before doing so, $d_f$ computes the following: a random key $t$; the hash values $r = H(t\, ||\, \pub{d_f})$ and $s = H(r)$; the encryption of the package $P_e = Enc(P, r)$. Now the proof $\pi$ is obtained using both the secret values ($P$ and $r$) and the public values for $S_E$ as inputs, along with $\pke$. This proof is sent to $d_s$ along with $P_e$, $s$, $\vke$ and $\pke$. The keys can be verified by $d_s$ using their hash values on the BC. Afterwards, $d_s$ verifies the proof $\pi$ using the public values $P_h$ (retrieved from the BC), $P_e$ and $s$, as well as $\vke$. If the proof is valid, $d_s$ is mathematically convinced that: (1) $P_e$ was effectively obtained encrypting a package $P$ with hash value $P_h$ using a key $r$; and (2) $s$ was indeed obtained computing the hash of $r$. At this point, $d_s$ can send a transaction (attaching an arbitrary CC amount as a offer) to the SSC to deploy a new ESC, the second kind of SC that can be generated. The code for an ESC is similar to the one for a DSC and can be summarized as follows: if the ESC is not expired, a CC payment will be issued to any sender with PK $\pub{d_f}$ who is able to provide the values $t$ and $r$ so that $r = H(t\, ||\, \pub{d_f})$ and $s = H(r)$; when these conditions are met, the key $r$ will also be published on the BC. Now $d_f$ can send $t$ and $r$ to the ESC, triggering the CC payment. The key $r$ can then be used by $d_s$ to decrypt $P_e$ and obtain $P$. From this moment on, there is no difference between $d_f$ and $d_s$, they both have all necessary elements to deliver the update to IoT targets and obtain CC rewards.

\subsubsection{Update Delivery} \label{delivery}

Let us continue with the actions of a distributor $d$, which can be either $d_f$ or $d_s$, or anyone in possession of the package $P$. The hub $h$, responsible for the IoT object $o_m$, discovers about a new update by looking at the BC and sends a request through the P2P network using the value $U_h$, attaching a fresh nonce $n_1$. The request is received by $d$, which sends back its PK $\pub{d}$, a signature $\textit{sig}_d := \sign{d}{n_1}$ and an ID challenge $c$. The hub verifies $\textit{sig}_d$ and, if valid, checks the integer score corresponding to $\pub{d}$ on the BC. If the latter is not satisfying, $h$ can search for another distributor. Otherwise, $c$ is forwarded by $h$ to $o_m$, which in turn generates a fresh nonce $n_2$ and replies with a signature $\textit{sig}^{\textit{ID}}_{o_m} = \sign{o_m}{c\, ||\, n_2}$, attaching also $n_2$. The hub forwards ${sig}^{\textit{ID}}_{o_m}$ and $n_2$ to $d$, along with the PK of the object $\pub{o_m}$, so that $d$ can (1) verify the signature ${sig}^{\textit{ID}}_{o_m}$ and (2) verify that $\pub{o_m}$ belongs to the list $L_m$ on the BC.

Now $d$ can prepare for the zk-SNARKs proof computing the following: a random key $t$; the hash values $r = H(t\, ||\, \pub{o_m}\, ||\, \pub{d})$ and $s = H(r)$; the encryption of the update file $U_e = Enc(U, r)$. The proof $\pi$ can then be obtained employing both the secret values ($U$ and $r$) and the public values for the statement $S_D$ as inputs, along with the proving key $\pkd$. Finally, $d$ can send $\pi$ to $h$, along with $U_e$, $s$, $\vkd$ and the signature $\textit{sig}_m$. The hub verifies the proof $\pi$ using the public values of $S_D$ and $\vkd$. If valid, $h$ is mathematically convinced that: (1) $U_e$ was indeed obtained encrypting an update file $U$ with hash value $U_h$ using a key $r$, where $U$ and $r$ are still unknown; (2) $s$ is the hash of $r$. Therefore, $h$ forwards $U_h$, $sig_m$ and $s$ to the IoT device $o_m$. In turn, $o_m$ verifies $sig_m$ and sends back a signature $\sigpod{PoD}{o_m} := \sign{o_m}{U_h\, ||\, s}$. This signature is the PoD that can be sent by $d$ to the DSC to unlock the CC reward. Once received by $h$, $\sigpod{PoD}{o_m}$ is forwarded to $d$.

\subsubsection{Key Publication and Final Delivery}

The distributor $d$ first verifies the signature $\sigpod{PoD}{o_m}$ against the public key $\pub{o_m}$ received before. If valid, $d$ can post a transaction to the DSC, attaching the following: (1) the PK of the targeted object $\pub{o_m}$; (2) the values $t$, $r$ and $s$; (3) the PoD $\sigpod{PoD}{o_m}$. If not yet expired, the DSC automatically checks the following: (1) the object corresponding to $\pub{o_m}$ is in the list and was not already served by another distributor; (2) the equalities $r = H(t \, || \, \pub{o_m} \, || \, \pub{\textit{sender}})$ and $s = H(r)$ are satisfied; (3) the signature $\sigpod{PoD}{o_m}$ is verified. If all checks are successful, a CC payment of amount $a_d$ is issued to $\pub{\textit{sender}}$, which is $\pub{d}$ in this case, and the key $r$ is published on the BC. Now $h$ can retrieve $r$ and decrypt $U_e$ to obtain $U$.
At this point, another important action is performed by the DSC. That is, incrementing the score of distributor $d$, associated with its public key. This is done automatically, sending a transaction to the parent contract, i.e. the SSC. The latter checks if this is the first delivery for this specific distributor: if yes, it creates a new entry for it, with value $1$. Otherwise, it checks if the reset period is expired: in this case the score is reset to $1$ for the existing entry. Finally, in the third case, the score is incremented. The integrity and security of these operations is enforced through internal BC mechanisms, as well as through the well-formedness of the code of both contracts.
%the journey of $U$ from the manufacturer $m$ to the IoT target $o_m$ is almost complete.
the hub sends $U$ to $o_m$, and expects the latter to send back a signature $\sigpod{PoFD}{o_m} := \sign{o_m}{U_h\, ||\, \pub{h}}$. Before doing so, $o_m$ computes the hash value of $U$, and compares it with the value $U_h$ received before, which was also authenticated by $m$ through the signature $sig_m$. The PoFD is then sent by $h$ to the DSC to obtain a CC reward. Before issuing the payment, the DSC checks that $o_m$ is in the list $L_m$ and was not yet served by another hub, and that the signature $\sigpod{PoFD}{o_m}$ is valid.

% \subfile{../msc/crowdp-msc.tex}

\section{Protocol Security} \label{security}
% We first present an informal analysis of the proposed protocol in Section \ref{informal}. Later in Section \ref{formal}, we provide the results of our formal analysis, applied to a simplified version of the protocol.

\subsection{Informal Analysis} \label{informal}

% \subsubsection{Replay Attack}

% We consider the case in which an adversary intercepts one or more messages exchanged in the protocol and retransmits them later to other (or the same) participants, in order to obtain a fraudulent result.
% In the very first stage of the \textit{CrowdPatching} protocol, \attacker could reply the messages sent by the manufacturer to deploy the SSC, or the DSC, to the BC. However, the only result would be to aid such deployment, since any transaction is secured through the use of cryptographic signatures. \mynote{[Replay attack is probably not relevant in this context]}.
% Referring to the steps in Section \ref{delivery}, the attacker could intercept the request for the update package sent by $h$ to the P2P network and send a duplicate later, when $h$ is already exchanging messages with $d$. However, this would bring no harmful consequences. If another distributor $d_1$ tries respond to the duplicate request, $h$ will not reply to the corresponding challenge, and $d_1$  will drop the connection. The adversary could also replay the challenge response sent by $h$ to $d$, redirecting it to $d_1$. However, we assume challenges to be fresh random values, which makes it impossible for the replayed response, originally meant for $d$, to be accepted by $d_1$.
% Another possibility for the attacker could be to intercept the zk-SNARK proof sent by $d$ to $h$, along with the additional values attached in the same message.

\subsubsection{Impersonation Attacks}

An attacker $\alpha$ could try to masquerade as a manufacturer $m$, deploying a new SSC and then a DSC, with the intent of delivering a malicious update to certain IoT devices. However, $\alpha$ would need to provide a CC fund to reward distributors and hubs. Additionally, due to the BC implementation, it would not be possible to impersonate $m$ without knowing its private key. Furthermore, the protocol requires $m$ to produce a signature on the hash value of the update, which is then validated by an IoT object offline. On the other hand, impersonating a distributor is indeed possible for $\alpha$. However, the protocol is intentionally designed to allow anyone to participate as a distributor. It is instead impossible for $\alpha$ to masquerade as a hub. We assume these entities to be manually selected, through secure user configuration, to manage one or more IoT devices. Therefore, an attacker without such privileges could neither (1) reply to any distributor asking for an identification challenge response, which must be produced by an IoT device, or (2) interact with any IoT device. Finally, it is not possible to impersonate an IoT device without the knowledge of its private key.

\subsubsection{PoD Submission Interception}

An attacker could intercept a transaction sent by $d$ containing a PoD submission, which includes the value $\sigpod{PoD}{o_m} := \sign{o_m}{U_h\, ||\, s}$, as well as $t$, $r$, $s$ and $\pub{o_m}$. The attacker could then replay some of these elements to the DSC, in order to obtain the CC reward in place of $d$. Before issuing the payment, the DSC checks the equations $r = H(t \, || \, \pub{o_m} \, || \, \pub{\textit{sender}})$ and $s = H(r)$. The attacker can easily generate a new value for $t$, and then craft $r$ and $s$ in such a way that the equations would hold. However, the DSC would also verify the signature $\sigpod{PoD}{o_m}$, and the latter cannot be crafted by the attacker to match $s$.

\subsubsection{Malicious Distributor}

We consider the case of an attacker $\alpha$ acting as an honest distributor, obtaining the update file from the manufacturer and interacting with hubs to waste their resources. The attacker could produce a valid zk-SNARKs proof, without eventually submitting the PoD to the DSC. In this case, a hub would not be able to decrypt the received file. This attack is mitigated in two ways. Firstly, $\alpha$ would be strongly encouraged to submit the PoD, because of the CC payment that would follow. Furthermore, before committing any resources, hubs can easily check $\alpha$'s score on the BC. If too low, hubs can simply drop the connection and look for more trustworthy distributors in the P2P network.

\subsubsection{Update Integrity}

We consider the case in which an attacker $\alpha$ attempts to deliver a modified update, with the goal of inducing an IoT device to execute malicious code. This is impossible for many reasons. First of all, while impersonating a honest distributor, $\alpha$ would need to generate a zk-SNARKs proof. In order to be valid, the proof must be generated encrypting an update file matching the hash value published on the BC by the manufacturer. The latter can be consulted by a hub at any time. As a consequence, a honest hub will never accept an unauthenticated file, unless the BC or the zk-SNARKs system are not secure. The only other way for $\alpha$ to successfully deliver a malicious update is to compromise a hub gateway. However, in this case the protocol remains secure. Before signing the identification challenge, an IoT device must verify a signature made by the manufacturer on the hash value of the new update. Later, in the very last step of the protocol, upon receiving the update file, the IoT device will check if it matches the authenticated hash value. In this way, it is impossible for such devices to accept a malicious update, unless (1) the device itself is compromised or (2) the private key of the manufacturer is stolen by $\alpha$.

\subsubsection{Dropping BC Transactions}

Various stages of the protocol are based on entities submitting transactions to the BC. We consider the case of an attacker $\alpha$ attempting to intercept these transactions, preventing them from reaching the BC network. To achieve this, $\alpha$ would need to block all connections from the targeted node. Given the distributed nature of such network, we argue that this would be unfeasible.

\subsection{Formal Analysis with Tamarin} \label{formal}

The Tamarin Prover is a protocol verification tool supporting automated analysis in the symbolic model. It has been applied to several real-world protocols, such as TLS 1.3 \cite{cremers_comprehensive_2017}, to verify complex security properties. 
The attacker is modeled as a Dolev-Yao adversary, controlling the entire network but incapable of breaking cryptographic functions.
A protocol is described through multiset rewrite rules and facts. Facts are used to model all the possible components of the system states, while rules describe the transitions between those states. A fact $F$ with arity $n$ is defined as $F(t_1, t_2, ... , t_n)$, where each $t_i$ is a term. Rules are composed by an initial state in the left-hand side and a final state in the right-hand side, each containing an arbitrary number of facts. Optionally, rules can have transition labels, called action facts.
A rule with initial state \lstinline{X}, final state \lstinline{Y} and a set of actions \lstinline{Z}, is formalized as: \lstinline{[X] --[Z]-> [Y]}. 
After having defined a certain model, one can specify its security properties, called lemmas.
% They are defined using formulas in a first-order logic with timepoints.
Tamarin checks that a property holds exploring all possible protocol executions. It either returns a proof that the property holds, or a counterexample representing an attack. However, the verification could also never terminate.

% \mynote{Tamarin supports C-style comments.}

In the symbolic model employed by Tamarin, the properties of the cryptographic functions are specified with equations over fact terms. For example, the built-in  symmetric encryption is defined through the functions \lstinline{enc} and \lstinline{dec}, plus the equation \lstinline{dec(enc(m,k),k) = m}, where \lstinline{m} is the plaintext and \lstinline{k} the secret key. Many other built-in primitives are available. In our model, we introduce custom definitions for the zk-SNARKs proving system. Two constant functions are used to model the generation of the proving key and the verifying key. We declare them as private so that the attacker cannot execute them, as these keys are generated securely by the manufacturer.
% However, we allow the possibility for an attacker to obtain them once they are generated.
We also define two functions corresponding to the \textit{Prove} and \textit{Verify} algorithms.
% The former takes two parameters: (1) the proving key and (2) the set of all secret and non-secret values. The latter takes three: (1) the verifying key, (2) the set of non-secret parameters and (3) the proof to be verified.
Furthermore, another constant function represents the output value of \textit{Verify} when the proof is valid. Finally, we introduce an equation linking all the previous functions to specify the case in which a proof is valid:

\begin{lstlisting}
functions:
	GenProvKey/0 [private], GenVerifKey/0 [private],
	zkProve/2, zkVerify/3, ver/0
equations:
	zkVerify( GenVerifKey, <h(U), senc(U, r), h(r)>,
		zkProve(GenProvKey,
			<<h(U), senc(U, r), h(r)>, <U, r>>) ) = ver
\end{lstlisting}

We model a simplified version of our protocol in which a single manufacturer \lstinline{M} releases an update for three IoT devices \lstinline{IoT1}, \lstinline{IoT2} and \lstinline{IoT3}. Two of this devices are managed by one hub, \lstinline{H1}, the other by \lstinline{H2}. The number of distributors is limited to three: \lstinline{D1}, \lstinline{D2} and \lstinline{D3}. We avoid modeling the SSC, and we directly generate the DSC instead. We also omit the initial seeding phase, thus assuming that distributors are already in possession of the update file, as well as hub rewards. All these entities are initialized through a setup rule. We also model the BC through specific persistent facts. In general, persistent facts can be consumed indefinitely, as opposed to default facts, called linear, which can be consumed only once. This persistency is useful to model immutable data on the BC. However, facts cannot be read by the attacker if not explicitly specified, while the BC should be accessible to anyone. For this reason, whenever a fact is used to represent data on the BC, the same data is also sent to the attacker-controlled network.
We report the setup rule in the following. In the left-hand side, private keys are instantiated for all entities, and the new update file is created through the built-in fact \lstinline{Fr}, which provides fresh values.
Facts describing public/private key pairs were generated (by a previous rule) as suggested by the official Tamarin manual\footnote{https://tamarin-prover.github.io/manual/}.
In the right-hand side, initial states are created for each entity with the same template. For example, for \lstinline{IoT1}, a new fact \lstinline{St_IoT_0('IoT1', < ... >)} is created, where all needed terms are enclosed in triangular brackets. In this case, \lstinline{IoT1} needs to have knowledge about its own private key, the public key of \lstinline{M} and information on the corresponding hub. Analogous facts are created for the remaining devices, as well as hubs and distributors. However, each distributor starts with three different state-facts, one for each target IoT object, to represent their competition towards the same goal of serving all devices. We also create the initial state for the DSC, as well as facts for the public information on the BC. Finally, we also send all public information on the network through the built-in fact \lstinline{Out()}.
% \mynote{Two transition labels are indicated, which will be used to define security properties later.}

\begin{lstlisting}
rule setup:
let
	Uh=h(~U) PK=GenProvKey VK=GenVerifKey
	sigByM=sign(Uh, ~ltkM) P=<~U, PK, PH, sigByM> 
in
[ Fr(~U), !Ltk('M', ~ltkM), !Ltk('IoT1', ~ltkIoT1)
, ... , !Ltk('D1',~ltkD1), ... , !Ltk('H2', ~ltkH2)]

--[Setup(), UpdatePublished(~U)]->

[ St_IoT_0('IoT1', <~ltkIoT1, pk(~ltkM), 'H1'>), ...
, St_D_0( 'D1', <~ltkD1, P, 'IoT1'> ), ...
, St_SC_0( 'SC', <'IoT1', pk(~ltkIoT1)> ), ...
, !DSC_Info_IoT( 'IoT1', pk(~ltkIoT1) ), ...
, Out( <Uid, 'IoT1', pk(~ltkIoT1)> ), ...
, !DSC_Info(Uh, h(VK)), Out(<Uh, h(VK), sigByM>) ]
\end{lstlisting}

After the setup rule, each protocol step is encoded with the same rule template. For example, the first rule represents a hub discovering about a new release from the DSC and requesting the update on the P2P network, where the hub transitions from state $0$ (output of the setup rule) to state $1$:

\begin{lstlisting}
rule H_1:
[ St_H_0( $H, <~ltkH, $IoT> )
, !DSC_Info( Uh, VKh, sigByM ) ]
--[ ]->
[ Out( <'UpdateRequest', $IoT, $H> )
, St_H_1( $H, <~ltkH, $IoT, Uid, sigByM, VKh> ) ]
\end{lstlisting}

Reporting all the rules describing the whole protocol is unfeasible: they can be found on the project GitHub link at \url{https://github.com/edoardopuggioni/crowdpatching}. Here, we will also report another significant example, the generation of the zk-SNARKs proof by a distributor. 

The rule starts with a message from a hub. The latter contains the signature on the ID challenge. This signature is verified by the distributor through an \lstinline{Eq} action fact. In the right-hand side, the distributor sends the proof to the hub and transitions to its next state-fact.

\begin{lstlisting}
rule D_2: let ... pi=zkProve(PK, <pub, sec>) ... in
[ St_D_1($D, <~ltkD, P, $IoT, pkIoT, $H, ~c>), ...
, In(<'IdReply', $H, $D, $IoT, sigCN, n>), Fr(~t) ]

--[Eq(verify, true), GenProof(pk(~ltkD), $IoT, U)]->

[ St_D_2($D, <~ltkD, P, $IoT, pkIoT, $H, ~t, r, s>)
, Out(<'zkProof', $D, $H, $IoT, proof, Ue, s, VK>) ]
\end{lstlisting}

We verified a series of security properties for this model. First of all, we demonstrated the executability of the protocol, ensuring that all rules are well-formed.
% The latter is a fundamental result, without which one cannot proceed to prove any other property. In few words, it ensures that all protocol rules are well-formed. 
Additionally, we managed to verify three important lemmas. The first verifies that a distributor receives a payment only if a valid zk-SNARKs proof was delivered, along with the encrypted update. We executed this lemma with the automatic mode provided by Tamarin, and the verification process terminated successfully. This means that the property holds for all possible executions of the protocol. In general, lemmas are defined as mathematical properties, referring to timepoints through the symbols \lstinline{#} and \lstinline{@}, and to action facts through their labels. In this case, the lemma says that for all distributors receiving an update, it must be true that the same distributor produced a zk-SNARKs proof:

\begin{lstlisting}
lemma PaymentOnlyIfGenerateProof: all-traces
"All #j pkD IoT . PaymentToD(pkD, IoT) @j ==>
 Ex #i U . GenProof(pkD, IoT, U) @i & i<j"
\end{lstlisting}

In a similar way, we also proved that (1) a distributor always receives a CC payment if the update is delivered to an IoT device and (2) it is not possible to receive more than one payment for a single IoT device.
The lemmas encoding these properties are defined as follows:

\begin{lstlisting}
lemma AlwaysPaidIfUpdateReady: all-traces
"All #k IoT U . UpdateReadyForIoT(IoT, U) @k
==> Ex #i #j pkD. PaymentToD(pkD, IoT) @i
	  & GenProof(pkD, IoT, U) @j"

lemma MaxOnePaymentForOneIoT: all-traces
"All #i #j IoT pkD1 pkD2 . PaymentToD(pkD1, IoT) @i 
   & PaymentToD(pkD2, IoT) @j ==> #i=#j & pkD1=pkD2"
\end{lstlisting}
% \section{Implementation}

\section{Conclusions} \label{concl}
We proposed a blockchain-based decentralized protocol, allowing manufacturers to delegate the delivery of software updates to self-interested distributors in exchange for cryptocurrency. We introduced significant improvements with respect to the most recent proposals addressing key limitations with respect to scalability, practicality, and security. In order to ensure about security of CrowdPatching we informally analyzed the most significant threats applicable to our protocol. Furthermore, compared with related work, we also performed a formal analysis by means of the Tamarin Prover to provide reliant assurance on the security of the protocol as well as its correctness. In our future work, we intend to develop a full prototype implementation of the protocol and extensively evaluate its performance in real deployments.

% \section*{Acknowledgment}

\bibliography{biblio}
\bibliographystyle{IEEEtran}

\end{document}

%% file: crowdp-msc.tex
% \begin{figure*}[htb!]
\begin{figure*}[!t]
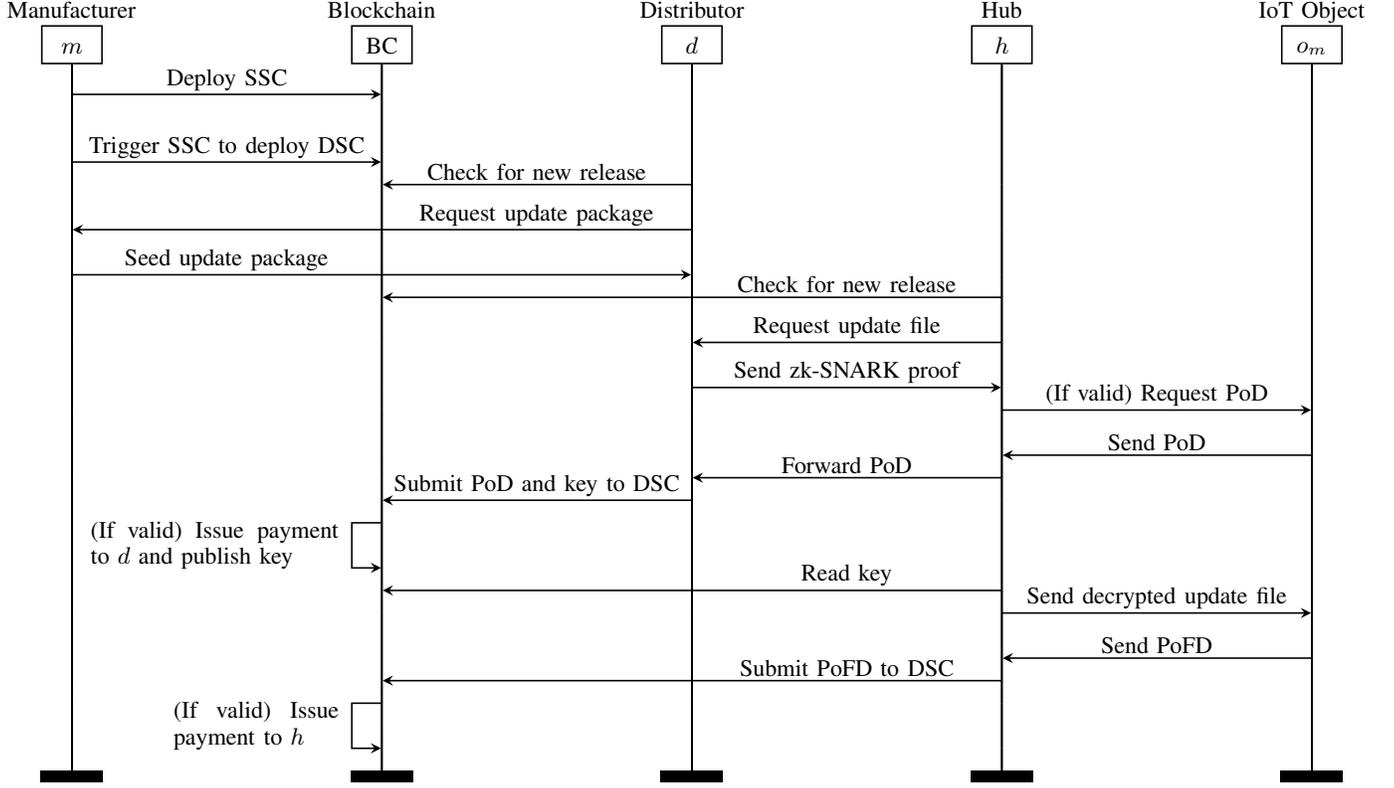

\centering

    \setmsckeyword{}
    \drawframe{no}
    
    \begin{msc}
        [ small values % small elements
        , /msc/environment distance=0.0\mscunit % no margins
        , /msc/msc keyword= % no header
        %, /msc/label distance=.3ex % distance label-arrow (messages)
        , /msc/level height=0.3cm
        ]{}

    % Alternative lengths
    \setlength{\instwidth}{0.8\mscunit}
    \setlength{\instdist}{3.3\mscunit}
    
    \declinst{m}{Manufacturer}{$m$}
    \declinst{bc}{Blockchain}{BC}
    \declinst{d}{Distributor}{$d$}
    \declinst{h}{Hub}{$h$}
    \declinst{iot}{IoT Object}{$o_m$}
    
    % \msccomment{Example comment}{m}

    % \nextlevel % add one level
    % \nextlevel[5] % add 5 levels
    
    % \action*[align=left]{Action line 1 \\ Action line 2}{m}

    % \mess[align=left, side=right, pos=.2]
    %     {
    %         Message to oneself
    %     }
    %     {m}{m}\

    % \mess[pos=.5]{\parbox{1cm}{\centering two\\lines}}{m}{bc}
    % \nextlevel[2]

    % Relative positions for message labels:
    % 0 entities in between: 0.50
    % 1 entities in between: 0.25
    % etc.
    
    \mess[pos=.5, /msc/label distance=.3ex]
    {Deploy SSC}{m}{bc}
    \nextlevel[3]
    
    \mess[pos=.5, /msc/label distance=.3ex]
    {Trigger SSC to deploy DSC}{m}{bc}
    \nextlevel[1]
    
    \mess[pos=.5, /msc/label distance=.3ex]
    {Check for new release}{d}{bc}
    \nextlevel[2]
    
    \mess[pos=.25, /msc/label distance=.3ex]
    {Request update package}{d}{m}
    \nextlevel[2]
    
    \mess[pos=.25, /msc/label distance=.3ex]
    {Seed update package}{m}{d}
    \nextlevel[1]

    \mess[pos=.25, /msc/label distance=.3ex]
    {Check for new release}{h}{bc}
    \nextlevel[2]

    \mess[pos=.5, /msc/label distance=.3ex]
    {Request update file}{h}{d}
    \nextlevel[2]
    
    \mess[pos=.5, /msc/label distance=.3ex]
    {Send zk-SNARK proof}{d}{h}
    \nextlevel[1]

    \mess[pos=.5, /msc/label distance=.3ex]
    {(If valid) Request PoD}{h}{iot}
    \nextlevel[2]

    \mess[pos=.5, /msc/label distance=.3ex]
    {Send PoD}{iot}{h}
    \nextlevel[1]

    \mess[pos=.5, /msc/label distance=.3ex]
    {Forward PoD}{h}{d}
    \nextlevel[1]

    \mess[pos=.5, /msc/label distance=.3ex]
    {Submit PoD and key to DSC}{d}{bc}
    \nextlevel[1]

    \mess[align=center, side=left, pos=0.5, /msc/label distance=1.0ex]
    {\parbox{3.3cm}{(If valid) Issue payment to $d$ and publish key}}{bc}{bc}
    \nextlevel[3]

    \mess[pos=.25, /msc/label distance=.3ex]
    {Read key}{h}{bc}
    \nextlevel[1]

    \mess[pos=.5, /msc/label distance=.3ex]
    {Send decrypted update file}{h}{iot}
    \nextlevel[2]

    \mess[pos=.5, /msc/label distance=.3ex]
    {Send PoFD}{iot}{h}
    \nextlevel[1]

    \mess[pos=.25, /msc/label distance=.3ex]
    {Submit PoFD to DSC}{h}{bc}
    \nextlevel[1]

    \mess[align=center, side=left, pos=0.5, /msc/label distance=1.0ex]
    {\parbox{2.2cm}{(If valid) Issue payment to $h$}}{bc}{bc}
    \nextlevel[1]

    \end{msc}
    
\caption{Overview of the protocol omitting DDE}
\label{fig:msc1}
\end{figure*}